\documentclass[12pt,preprint]{aastex}

\newcommand{\Rmnum}[1]{\expandafter\@slowromancap\romannumeral #1@}

\usepackage{natbib}
\shorttitle{On the enhanced CME rate since the cycle 23 polar reversal}
\shortauthors{Petrie}

\begin{document}


\title{On the enhanced coronal mass ejection detection rate since the solar cycle 23 polar field reversal}


\author{G.J.D. Petrie}
\affil{National Solar Observatory, Tucson, AZ 85719, USA}



\begin{abstract}
Coronal mass ejections (CMEs) with angular width $> 30^{\circ}$ have been observed to occur at a higher rate during solar cycle 24 compared to cycle 23, per sunspot number. This result is supported by data from three independent databases constructed using Large Angle and Spectrometric Coronagraph Experiment (LASCO) coronagraph images, two employing automated detection techniques and one compiled manually by human observers. According to the two databases that cover a larger field of view, the enhanced CME rate actually began shortly after the cycle 23 polar field reversal, in 2004, when the polar fields returned with a 40\% reduction in strength and interplanetary radial magnetic field became $\approx 30\%$ weaker. This result is consistent with the link between anomalous CME expansion and heliospheric total pressure decrease recently reported by Gopalswamy et al.
\end{abstract}



\section{Introduction}
\label{sect:intro}

Coronal mass ejections (CMEs) are large, twisted magnetic structures that are expelled from the solar surface out into the heliosphere at speeds of hundreds of km~s$^{-1}$. They are believed to remove magnetic helicity and free magnetic energy from the corona that would otherwise be difficult to eliminate, owing to the high electrical conductivity of the corona \citep{Low01}. They cause the most dangerous space weather effects on Earth \citep{WebbHoward12}. First observed in coronagraph images in the 1970s \citep{Tousey73}, they have since been observed systematically by numerous observatories \citep{WebbHoward94}, including since 1996 the Large Angle and Spectrometric Coronagraph Experiment (LASCO) \citep{Brueckneretal95} on NASA's Solar and Heliospheric Observatory (SoHO) satellite.

The relationship between the behavior of coronal mass ejections (CMEs) and the global solar magnetic field has long been a topic of lively interest. \citet{Luhmannetal98} and \citet{Lietal01} compared the locations of front-side CMEs observed by LASCO with the distributions of newly open magnetic field lines in the corona, inferring the opening field lines from potential-field source-surface models based on magnetogram observations taken before and after each CME. In some cases they found an intriguing morphological similarity between the CME emission structure and the newly-open field lines. Using LASCO data, \citet{Gopalswamyetal03b,Gopalswamyetal12} have reported an equatorward latitude offset between CME trajectories and the locations of their associated prominence eruptions, implying influence on the eruptions from the expanding polar coronal hole fields.
Furthermore, \citet{Gopalswamyetal03a} found a close relationship at both poles between the cycle 23 magnetic polarity reversals and the statistics of high-latitude CMEs: the high-latitude eruptions stopped in November 2000 and May 2002 in the northern and southern hemispheres, respectively, roughly coinciding with the north and south polar field reversals.

Several years ago it became clear that the cycle 23/24 minimum polar fields were approximately 40\% weaker than the cycle 22/23 minimum polar fields, accompanied by a 40-45\% decrease in polar coronal hole area \citep{Wangetal09,DeToma11}, and a 30\% decrease in the radial interplanetary mean field \citep[IMF][]{SmithBalogh08}. \citet{Luhmannetal11} suggested that the observed enhanced CME rate of cycle 24 may be explained in terms of the weak polar fields allowing weaker active region and prominence fields to erupt, and more ejections to escape into the heliosphere. \citet{Petrie13} found evidence in the Computer Aided CME Tracking (CACTus) and Solar Eruptive Events Data System (SEEDS), CME databases that the enhanced CME rates indeed correlated well in time with the weakening of the polar fields. In earlier data collected during cycle 21 and the rise of cycle 22 (1975-1989), the CME rate was very well correlated with the sunspot number \citep{WebbHoward94}. The LASCO CME rates based on the CACTus and SEEDS databases, in contrast, were systematically higher per sunspot number after the cycle 23 polar reversal compared to before, and the polar fields decreased in strength by about 40\% during that reversal \citep{Petrie13}. \citet{WangColaninno14} pointed out that the change in LASCO image cadence beginning in 2010, when the image rate approximately doubled, may have caused the increased rate of CME detections. Assuming that the CME detection rate is proportional to the LASCO image rate, they normalized the CME detection rate and found a much higher correlation with the sunspot number, and concluded that the polar fields have no significant affect on the occurrence rate of CMEs.

Since then, some evidence has appeared that the cycle 24 CME rate is genuinely enhanced over the cycle 23 rate, and a physical explanation has been offered. According to the Coordinated Data Analysis Workshop (CDAW) database, the halo CME rate has been higher during cycle 24 than during cycle 23 \citep{Gopalswamyetal15}, even though the sunspot number has been smaller by around 40\%. The distribution of CME sources in apparent longitude has also been much flatter, with proportionally twice as many halo CMEs originating from central meridian distances $\ge 60^{\circ}$. However, the average CME speed and flare size have been unchanged compared to cycle 23, leading \citet{Gopalswamyetal15} to suggest an explanation based on the ambient medium and not source magnetic fields: a decrease in total (magnetic + plasma) pressure in the corona and heliosphere.  \citet{Gopalswamyetal14} found evidence for this in a study of CME widths and velocities: the linear correlation between these two quantities has changed little from cycle 23 (Pearson linear correlation coefficient $r=0.67$ to cycle 24 ($r=0.72$), but the constant of proportionality has changed by 50\%. Cycle 24 CMEs are significantly wider than their cycle 23 counterparts without being significantly faster. Based on NASA OMNI solar wind data taken in situ at 1~AU, and assuming that the magnetic field strength, proton density and proton temperature decay with radial distance $R$ from the Sun as $R^{-2}$, $R^{-2}$ and $R^{-0.7}$, respectively, the authors argued that the heliospheric total pressure has declined by around 40\% between cycles 23 and 24, allowing the CMEs to expand more in the heliosphere, in turn allowing more CME detections to take place. This can also explain the reduced geoeffectiveness of the CMEs. They have diluted magnetic energy content because of their greater expansion, and they interact with weaker ambient heliospheric fields: \citet{SmithBalogh08} earlier found that the radial interplanetary magnetic field component measured at high latitudes by Ulysses was about a third weaker during the spacecraft's third orbit (during cycle 23/24 minimum) than compared to its first orbit (during cycle 22/23 minimum). They also found agreement between the Ulysees results and NASA OMNI solar wind data taken at 1~AU; the radial interplanetary field component multiplied by the square of the radial distance from the Sun is independent of position \citep{Baloghetal95,Smithetal01}, so that {\it in situ} Ulysses and near-earth measurements can be compared over long periods of time. The dynamical and thermal pressures were also significantly smaller \citep[by 22\% and 25\%, respectively][]{McComasetal08}.

This paper will compare the CME rate statistics from the three databases, SEEDS, CACTus and CDAW, and search for evidence of real CME rate changes since cycle 23, and relate these changes to decreases in the solar and heliospheric field strength. In Section~\ref{sect:data} the data sets will be introduced. The data will be analyzed and the results presented in Section~\ref{sect:cmes}, and we will conclude in Section~\ref{sect:conclusion}.

\section{Data}
\label{sect:data}

LASCO has completed 18 years of nearly continuous white-light imaging of the K-corona, covering cycles 23 and 24. LASCO initially had three coronagraphs with overlapping fields of view, C1 (1.1-3~$R_s$), C2 (2-6~$R_s$), and C3 (3.7-32~$R_s$), where $R_s$ is the solar radius. The C1 camera did not survive the temporary loss of the SoHO spacecraft in 1998. CMEs are therefore identified using images from the C2 and/or C3 coronagraphs. Three online databases, the Coordinated Data Analysis Workshop\footnote{http://cdaw.gsfc.nasa.gov} \citep[CDAW,][]{Gopalswamyetal09}, the Solar Eruptive Events Data System\footnote{http://spaceweather.gmu.edu/seeds/lasco.php} \citep[SEEDS,][]{Olmedoetal08}, and the Computer Aided CME Tracking\footnote{http://sidc.oma.be/cactus/} \citep[CACTus,][]{Robbrechtetal09}, have been recording CMEs covering the LASCO era (1997-present). The CDAW CME identifications have been achieved by visual inspection of the LASCO C2 and C3 coronagraph images, whereas SEEDS and CACTus apply automated algorithms to identify CMEs without human intervention, SEEDS using C2 images and CACTus using images from both C2 and C3.

The SEEDS detection algorithm is based on projecting two-dimensional images onto one dimension and searching for signatures of bright regions moving radially outward as observed in a running-difference time series \citep{Olmedoetal08}. The height, velocity, and acceleration of the CME are automatically determined. The SEEDS database identifies CMEs from C2 images only. CACTus \citep{Robbrechtetal09} automatically detects CMEs in image sequences from LASCO C2 and C3 by constructing stackplots of C2 and C3 images, sharpened using the Hough transform, to detect motions of bright CME structures. Until 2010 the algorithm ran on level 0 images from the LASCO C2 and C3 instruments and since 2010 quick-look images from these instruments have been used. The detection method and the database are described and analyzed in detail by \citet{Robbrechtetal09}.

The CDAW database is based on lists compiled by human observers who review near-real-time movies of C2 and C3 images. These observers need to judge what constitutes an individual CME in, e.g., a close sequence of eruptions and outflows. Qualitative descriptions are added to many of the records in the CDAW database, such as ``poor'', ``very poor'' and ``C2 only''. These comments help us to determine the effects of these questionable detections on the overall CDAW statistics.

We therefore have identifications from two automated (CACTus and SEEDS) and one manual database, with two databases (CDAW and CACTus) based on C2 and C3 images, and one (SEEDS) based on C2 images alone. Compared to databases assembled manually by human operators, automatic CME detections might be more objective as the detection criterion is written explicitly in a program, but the lack of manual filtering also means that the database needs to be treated with some caution. Qualitative comparison with the CDAW database is therefore a useful test of the automated algorithms' results. The automated databases also provide a check of the objectivity of the human observers' detections in the CDAW database.

Comparisons between CME databases have been conducted in the past. CACTus generally identifies more CMEs than CDAW, and the CACTus CME rate is better correlated with the sunspot cycle, though the CACTus CME rate lagged behind the sunspot cycle by 6-12 months during cycle 23 \citep{Robbrechtetal09}. The CDAW database is affected by observer bias. It has been compiled by several observers since 1997. After 2004 the CDAW detection rate increased significantly because the project began to include very narrow CMEs previously disregarded.

\citet{Boursieretal09} found good agreement between the SEEDS and Automatic Recognition of Transient Events  and Marseille Inventory from Synoptic maps (ARTEMIS) databases, perhaps reflecting the restriction of these databases to C2 images. (The ARTEMIS database compiled CME detections using LASCO C2 synoptic maps, but now appears to be offline.) The recorded SEEDS CME widths tend to be much narrower than the CDAW and CACTus widths \citep{Byrneetal09} because the SEEDS algorithm refers to C2 but not C3 images. The CACTus CME width distribution tends to be scale-invariant, in contrast to the CDAW width distribution which has a peak around $30^{\circ}$ \citep{Robbrechtetal09}. CACTus has recorded many more narrow CMEs than CDAW, whose human observers tend to miss many narrow CMEs during solar maximum. When the activity level is low, small, faint CMEs are easier to catch than during solar maximum. On the other hand, automated databases sometimes miss wide (e.g., halo) CMEs that are picked up by human observers \citep[e.g.][]{Gopalswamyetal10}. The observational definition of a CME is not uniquely defined and varies from database to database, and also within a single database in the case of CDAW.

In this paper we will revisit the issue of consistency between the three databases CACTus, CDAW and SEEDS, focusing on the agreement between them regarding the question of changes of CME detection rates between cycles 23 and 24. These three independent solar eruption databases will provide a useful profile of eruption rates during the LASCO era, that can be related to the solar magnetic field behavior during the same period of time.


\section{Comparison of CME eruption distributions and rates}
\label{sect:cmes}

\subsection{Velocity distributions}

\begin{figure}[h]
\begin{center}
\resizebox{0.9\hsize}{!}{\includegraphics*{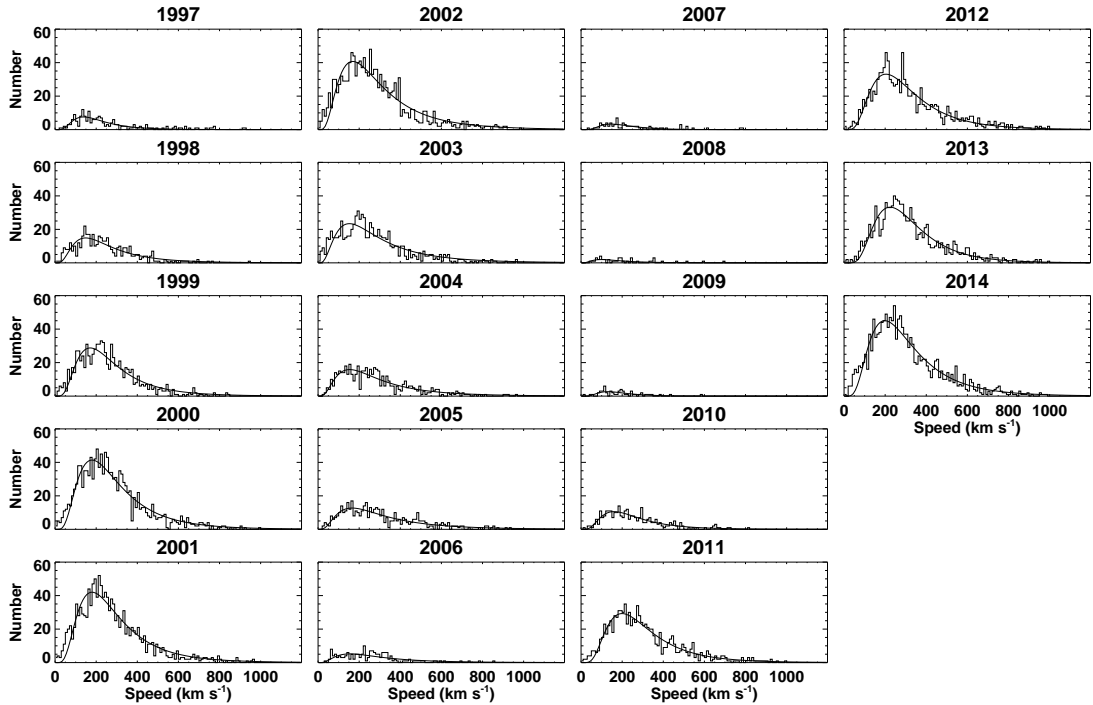}}
\end{center}
\caption{Annual histograms of CME velocity measurements from the SEEDS database. Fitted log-normal functions, described by Equation~(\ref{eq:lognormal}), are overplotted.}
\label{fig:velhists_seeds}
\end{figure}

\begin{figure}[h]
\begin{center}
\resizebox{0.9\hsize}{!}{\includegraphics*{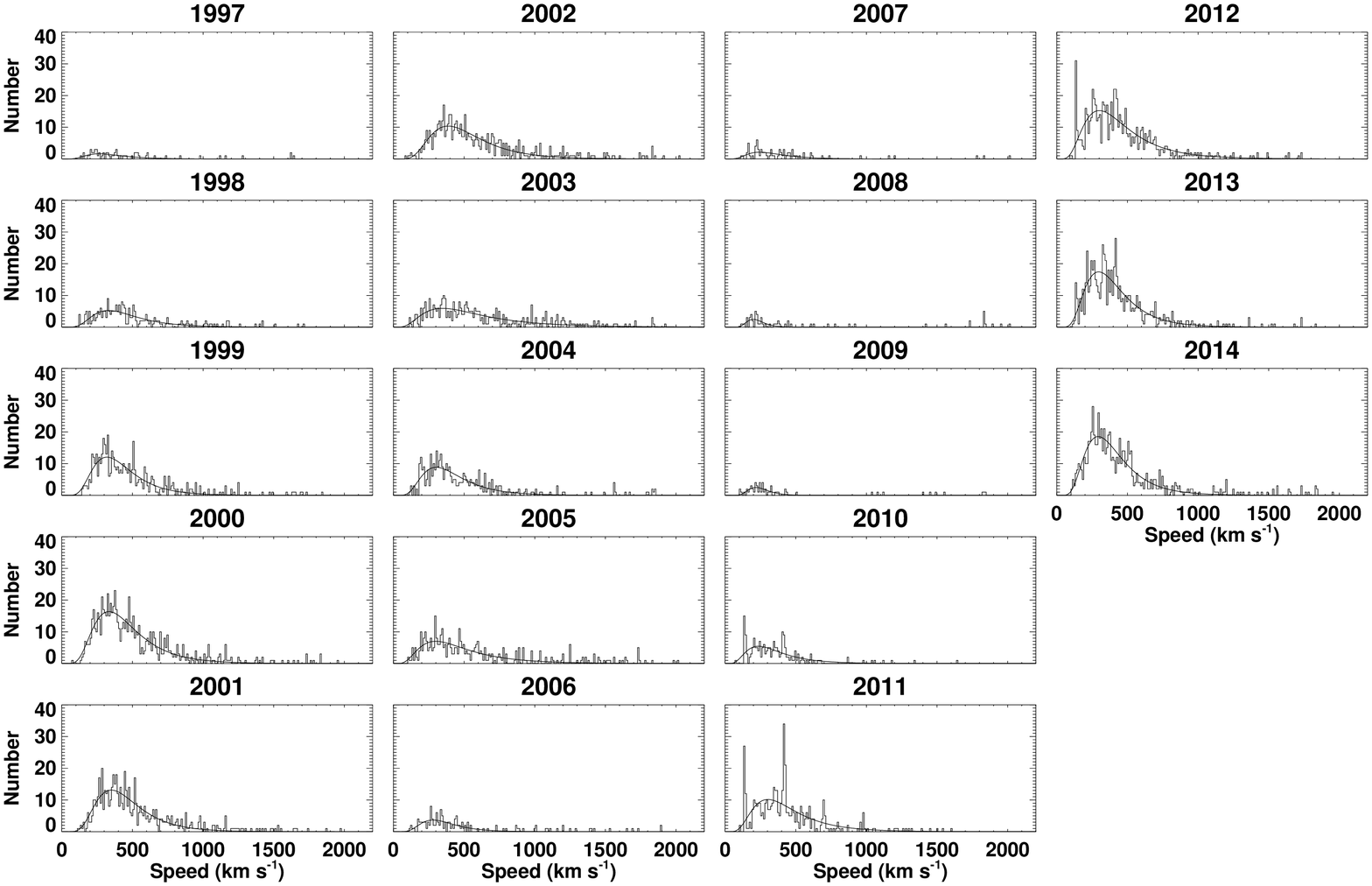}}
\end{center}
\caption{Annual histograms of CME velocity measurements from the CACTus database. Fitted log-normal functions, described by Equation~(\ref{eq:lognormal}), are overplotted.}
\label{fig:velhists_cactus}
\end{figure}

\begin{figure}[h]
\begin{center}
\resizebox{0.9\hsize}{!}{\includegraphics*{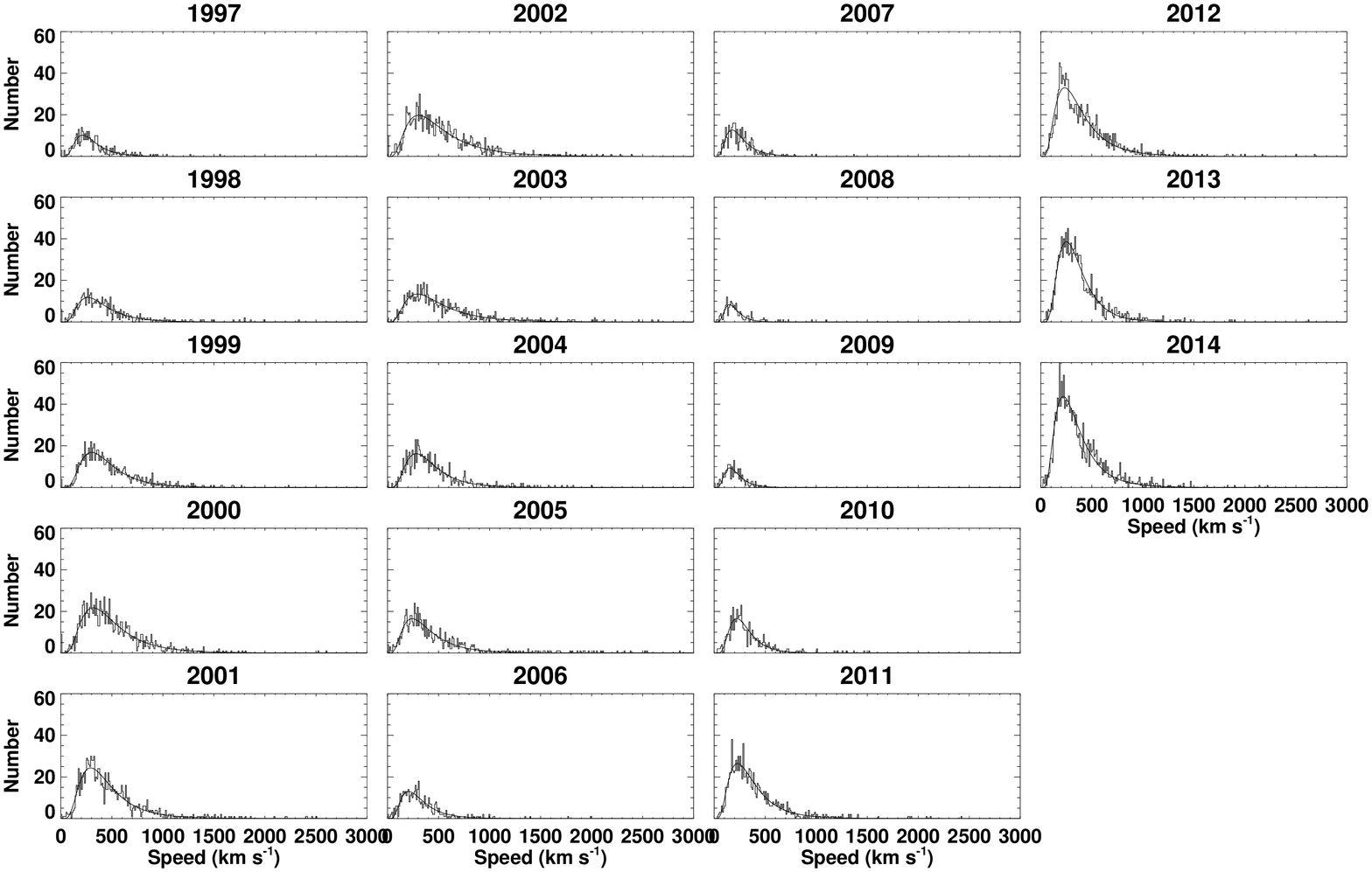}}
\end{center}
\caption{Annual histograms of CME velocity measurements from the CDAW database. Fitted log-normal functions, described by Equation~(\ref{eq:lognormal}), are overplotted.}
\label{fig:velhists_cdaw}
\end{figure}

\begin{figure}[h]
\begin{center}
\resizebox{0.9\hsize}{!}{\includegraphics*{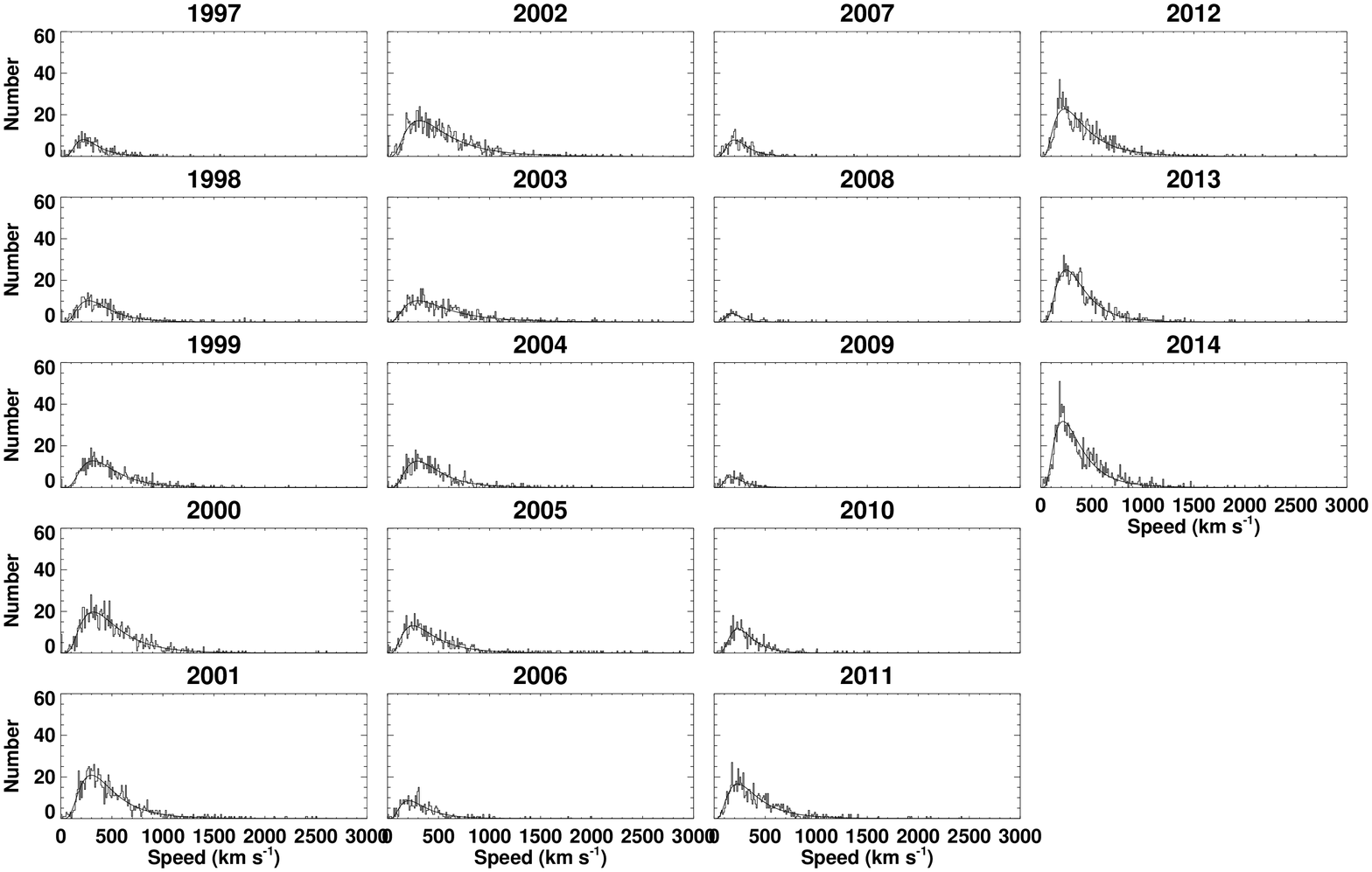}}
\end{center}
\caption{Annual histograms of CME velocity measurements from the CDAW database. Fitted log-normal functions, described by Equation~(\ref{eq:lognormal}), are overplotted. Poor and very poor detections have been excluded.}
\label{fig:velhists_cdawnop}
\end{figure}

Figures~\ref{fig:velhists_seeds}-\ref{fig:velhists_cdaw} show annual velocity histograms of detected CME velocities from the SEEDS, CACTus and CDAW databases, respectively. The histograms from all three databases all have approximately log-normal distributions, whose amplitudes have a clear solar cycle dependence. The log-normal distribution of CME speeds has been discussed in the past by, e.g., \citet{Yurchyshyn05}. Comparing the amplitudes of the log-normal distributions in Figures~\ref{fig:velhists_seeds}-\ref{fig:velhists_cdaw} year-by-year, the CACTUS detections seem to occur in consistently lower numbers than the SEEDS and CDAW detections. The CDAW distributions extend to higher velocities than the SEEDS and CACTUS distributions, though the CACTUS velocity histograms tend to peak around 300~km~s$^{-1}$, significantly higher than the SEEDS peak velocities and marginally higher than the CDAW peak velocities. On the other hand, the velocity distributions from all three databases appear to have similar widths. Figure~\ref{fig:velhists_cdawnop} shows the CDAW histograms with ``poor'' and ``very poor'' detections removed. The vast majority of detections survive, and Figures~\ref{fig:velhists_cdaw} and \ref{fig:velhists_cdawnop} appear similar, with slightly reduced statistics in Figure~\ref{fig:velhists_cdawnop}.

To gain a more quantitative characterization of the velocity distributions, we model the overall behavior of the velocity distributions year by year using the log-normal distribution function,

\begin{equation}
p(v)=\frac{C}{\sqrt{2\pi}\sigma v} e^{ \frac{-(\log{v} -\mu^2)}{2\sigma^2}}.
\label{eq:lognormal}
\end{equation}

\noindent This function describes a peaked, asymmetric distribution and is appropriate for modeling the velocity distributions shown in Figures~\ref{fig:velhists_seeds}-\ref{fig:velhists_cdawnop}. The location of the peak, referred to as the mode, is given by $e^{\mu-\sigma^2}$, and the median and mean by $e^{\mu}$ and $e^{\mu+{\sigma}/2}$. Thus the mode $\le$ the median $\le$ the mean, with equality only in the symmetric case $\sigma=0$. Overplotted on each histogram in Figures~\ref{fig:velhists_seeds}-\ref{fig:velhists_cdawnop} is the best log-normal fit to the histogram. The free parameters of the log-normal function in Equation~(\ref{eq:lognormal}) will be used later, in Subsection~\ref{subsect:lognormal}, to characterize and compare the velocity distributions. First we examine the effects of excluding classes of relatively unreliable detections.

\subsection{Effect of excluding narrow detections}

\begin{figure}[h]
\begin{center}
\resizebox{0.9\hsize}{!}{\includegraphics*{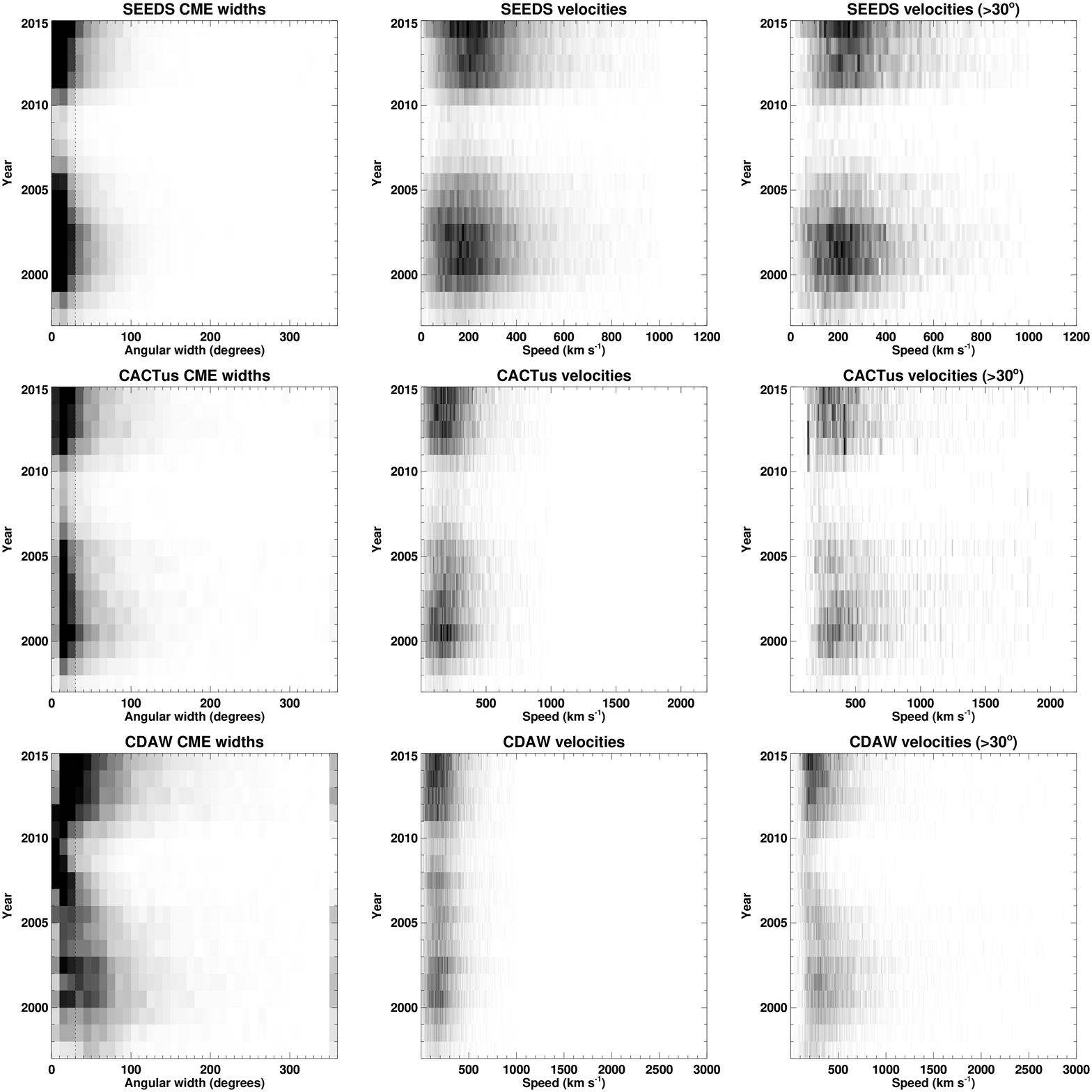}}
\end{center}
\caption{Stacked annual histograms of CME angular widths (left) and angular widths for all detections (middle) and for detections with angular width $> 30^{\circ}$ (right), from the SEEDS (top), CACTus (middle) and CDAW (bottom) databases. The dotted lines in the left plots indicate the angular width cutoff at $30^{\circ}$ below which the detections are excluded from the right plots.}
\label{fig:2dhists}
\end{figure}

Figure~\ref{fig:2dhists} shows the changes in time of the angular width (left column of plots) and velocity distributions (middle and right columns of plots) associated with the CME detections by SEEDS, CACTus and CDAW. The middle column of Figure~\ref{fig:2dhists} shows the velocity histograms for all CME detections and the right column shows the histograms with narrow cases (angular width $<30^{\circ}$) removed. The SEEDS data, based on C2 images only, clearly include a larger proportion of narrow CMEs than the other two data sets, which are based on data from C2 and C3. This may be because CMEs tend to expand super-radially as the ambient field strength drops off faster than 1/(radius)$^2$, so that as they travel from the C2 field of view to the C3 field of view they tend to have increasing angular width. The narrow CMEs detected by CACTus show a strong cycle-dependence, albeit with a higher rate of detection since 2010. The CDAW distribution shows more of a stepwise increase in narrow CME detections in 2007, with a broad peak during the cycle 23 minimum in 2007-2009. The CACTus detection increase may be related to the LASCO C2 and C3 image cadence increases that occurred during 2010, whereas the CDAW peak is likely due to the inclusion of narrower CMEs in the database by the human observers after 2004.

The exclusion of narrow CMEs in the right column of Figure~\ref{fig:2dhists} has the effect of reducing the number of slow CMEs and moving the peaks of the log-normal distributions up the velocity scale. This effect is not surprising because it is well known that the angular widths and velocities of CMEs are moderately well correlated \citep{Gopalswamyetal14}. This step removes most of the questionable, borderline CME detections: those with angular width $>30^{\circ}$ are relatively unlikely to be time-dependent plasma outflows mistakenly identified as CMEs. The velocity distributions in the right column of Figure~\ref{fig:2dhists} are more regular than those in the middle column as a result.

\subsection{Impact of LASCO image cadence increase}

\begin{figure}[h]
\begin{center}
\resizebox{0.9\hsize}{!}{\includegraphics*{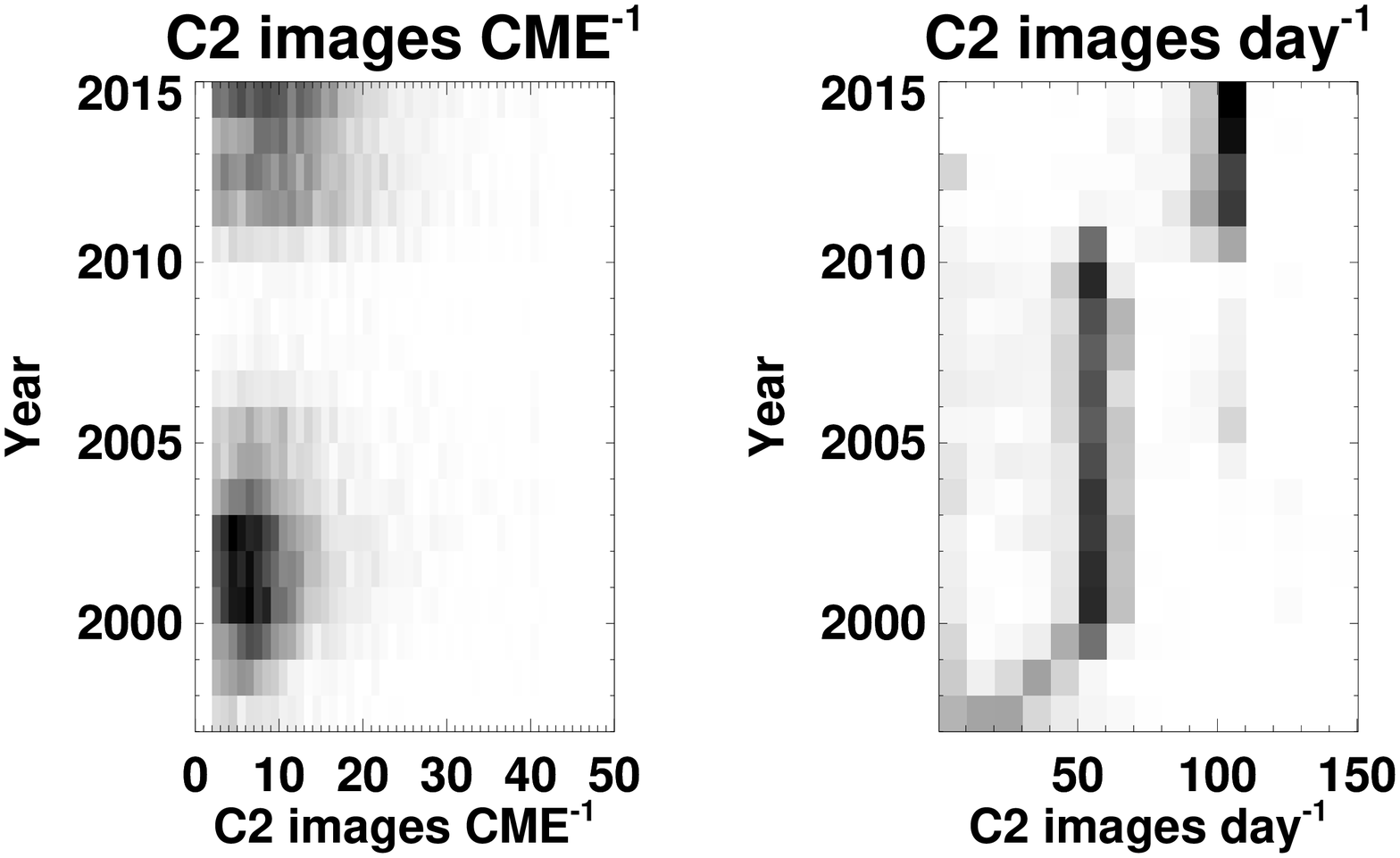}}
\end{center}
\caption{Stacked annual histograms of the number of LASCO C2 images per detected CME in the SEEDS database (left) and of the number of C2 images per day (right).}
\label{fig:nimhists}
\end{figure}

The cadence of LASCO C2 and C3 images increased by a factor of approximately two when other instruments on the Solar and Heliospheric Observatory (SoHO) ceased to send data to earth regularly, and extra bandwidth became available to LASCO. To assess the effect of the LASCO image cadence change on the rate of CME detection, Figure~\ref{fig:nimhists} shows a 2D histogram of annual number distributions of LASCO/C2 images per SEEDS CME detection and the average number of C2 images per day. Both statistics are available in the SEEDS database. The increase by a factor of about 2 of the nominal LASCO/C2 image cadence from about 50~day$^{-1}$ to about 100~day$^{-1}$ appears clearly in Figure~\ref{fig:nimhists} (right). This major change was accompanied by changes in the distribution of the number of C2 images per SEEDS CME detection, shown in Figure~\ref{fig:nimhists} (left). The figure shows that the distribution of images per CME detection broadened significantly when the C2 image cadence increased.

There is no obvious increase in the maximum speed of detected CMEs between cycles 23 and 24 in Figure~\ref{fig:2dhists}. According to Figure~\ref{fig:nimhists}, the main impact of the LASCO image cadence increase seems to have been to broaden the distribution of C2 images per SEEDS CME detection rather than introducing more detections of very fast or small, faint CMEs only detectable in a couple of high-cadence images. If the leading effect of the C2 image cadence increase had been to introduce many more detections then these would have included a large increase in detections in the left bins of the left plot of Figure~\ref{fig:nimhists}, either from very fast CMEs caught in just a few high-cadence images, or small, faint CMEs, caught in just a few images before becoming undetectably diffuse. However, these bins are less populated during cycle 24 than during cycle 23, a side-effect of the broadening of the distributions.

\subsection{Effect of excluding ``poor'' and ``very poor'' CDAW detections}

\begin{figure}[h]
\begin{center}
\resizebox{0.9\hsize}{!}{\includegraphics*{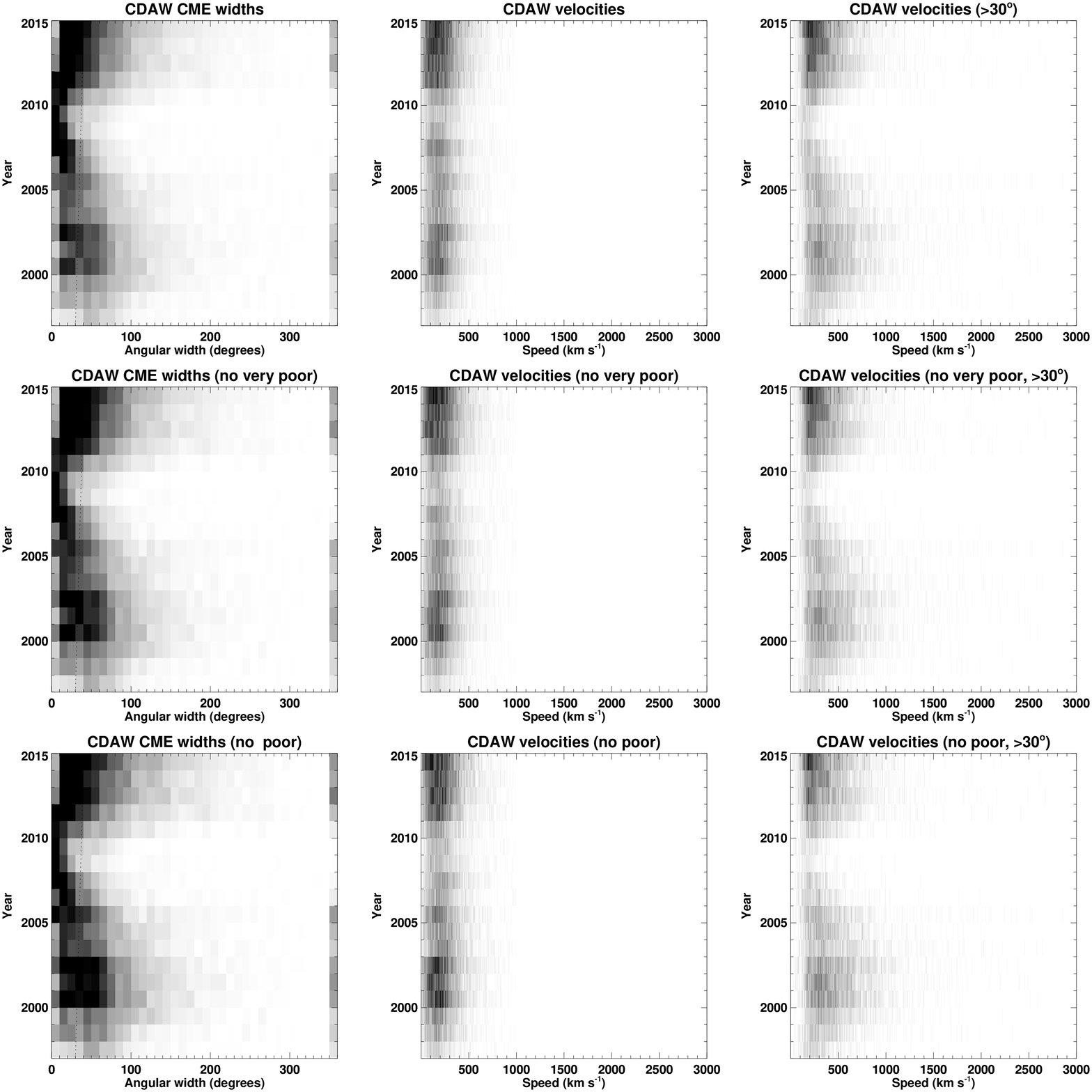}}
\end{center}
\caption{Stacked annual histograms of CME angular widths (left) and angular widths for all detections (middle) and for detections with angular width $> 30^{\circ}$ (right), from the CDAW database. All detections are included in the top plots, very poor detections are excluded from the middle plots, and poor and very poor detections are excluded from the bottom plots. The dotted lines in the left plots indicate the angular width cutoff at $30^{\circ}$ below which the detections are excluded from the right plots.}
\label{fig:cd2dhists}
\end{figure}

Figure~\ref{fig:cd2dhists} shows 2D histograms of the CDAW CME velocities and angular widths for the cases including all detections, and with ``poor'' and ``very poor'' detections removed. The right column of plots shows the velocity distributions for cases with angular width $> 30^{\circ}$. The top row of  Figure~\ref{fig:cd2dhists} is identical to the bottom row of Figure~\ref{fig:2dhists}. The curious maximum in the number of narrow CMEs coinciding with solar activity minimum (2007-2009) is again clearly visible in the left plots, including the plots with ``poor'' and ``very poor'' detections removed. To avoid artifacts associated with observer bias it is evidently not sufficient to remove the ``poor'' and ``very poor'' detections. Implausibly large post-2004 statistics are evident in the columns up to $30^{\circ}$ in all three left plots of Figure~\ref{fig:cd2dhists}. Beyond $30^{\circ}$ the histograms resemble the CACTus histogram in Figure~\ref{fig:2dhists} (middle left). The main difference between the left plots of Figure~\ref{fig:cd2dhists} for CMEs with angular width $>30^{\circ}$ is that the post-2004 statistics are clearly larger in the top left plot than the pre-2004 statistics because of the observer bias. This artifact is effectively removed by the exclusion of ``poor'' and ``very poor'' detections. We can verify this by qualitatively comparing the bottom left plot of Figure~\ref{fig:cd2dhists} to the middle left plot of Figure~\ref{fig:2dhists} which shows the distribution from the more objective CACTus algorithm processing the same C2 and C3 images.

The velocity distributions in the middle column of Figure~\ref{fig:cd2dhists} include detections of all angular widths. For comparison, the equivalent velocity distributions for detections with angular width $>30^{\circ}$ are plotted in the right column of Figure~\ref{fig:cd2dhists}. As in Figure~\ref{fig:2dhists}, the main effect of excluding narrow CMEs is to reduce the number of slow CMEs. Comparing the plots in the middle and right  columns of Figure~\ref{fig:cd2dhists}, removal of the ``poor'' and ``very poor'' detections seems to have only a minor effect on the velocity distribution, whereas removing narrow CMEs has a major simplifying effect, eliminating irregularities in the distribution and bringing it into much closer qualitative agreement with the CACTus distribution in Figure~\ref{fig:2dhists}.

Figures~\ref{fig:cd2dhists} and \ref{fig:2dhists} indicate that we can make an improved qualitative comparison between CACTus and CDAW statistics if we confine our attention to CMEs with angular width $>30^{\circ}$ and ignore ``poor'' and ``very poor'' CDAW detections. Figure~\ref{fig:cd2dhists} shows also that the statistics of halo CME detections (angular width 360$^{\circ}$) are not changed much by the exclusion of ``poor'' and ``very poor'' detections. Halo CMEs form a small but notably robust subset of the CDAW database \citep{Gopalswamyetal15}.

\subsection{Comparison of the log-normal distribution parameters in time}
\label{subsect:lognormal}

\begin{figure}[h]
\begin{center}
\resizebox{0.45\hsize}{!}{\includegraphics*{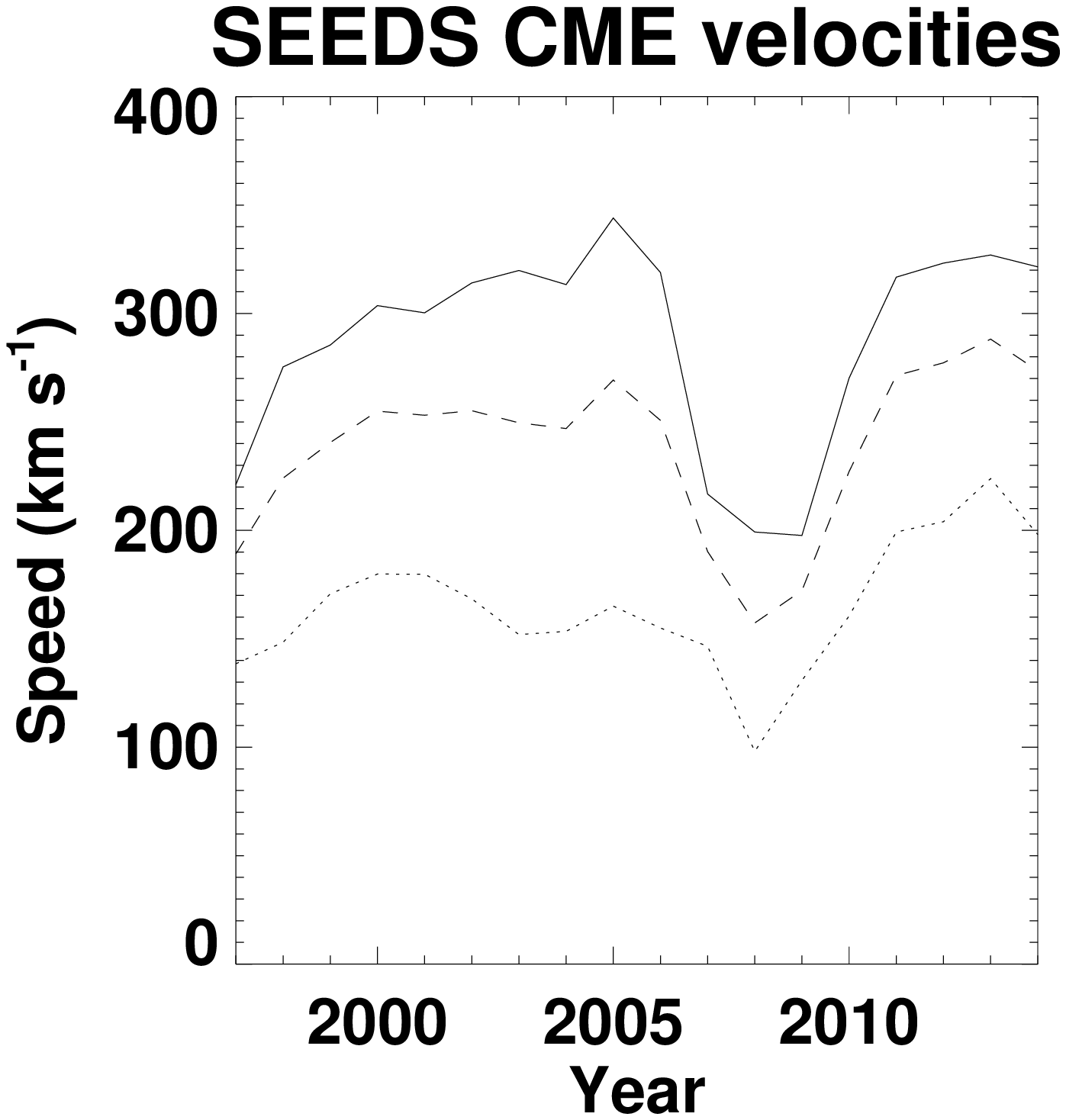}}
\resizebox{0.45\hsize}{!}{\includegraphics*{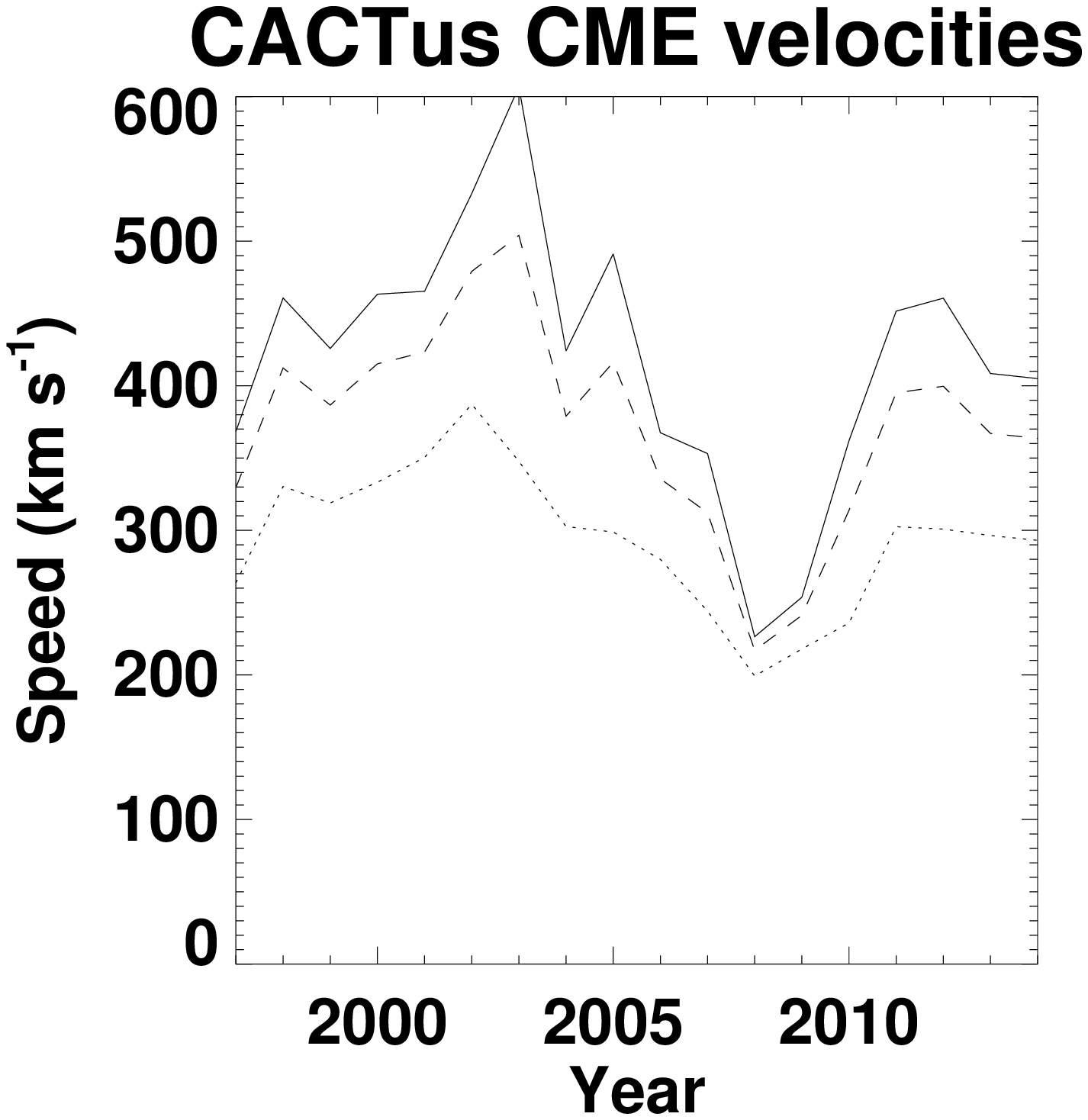}}
\resizebox{0.45\hsize}{!}{\includegraphics*{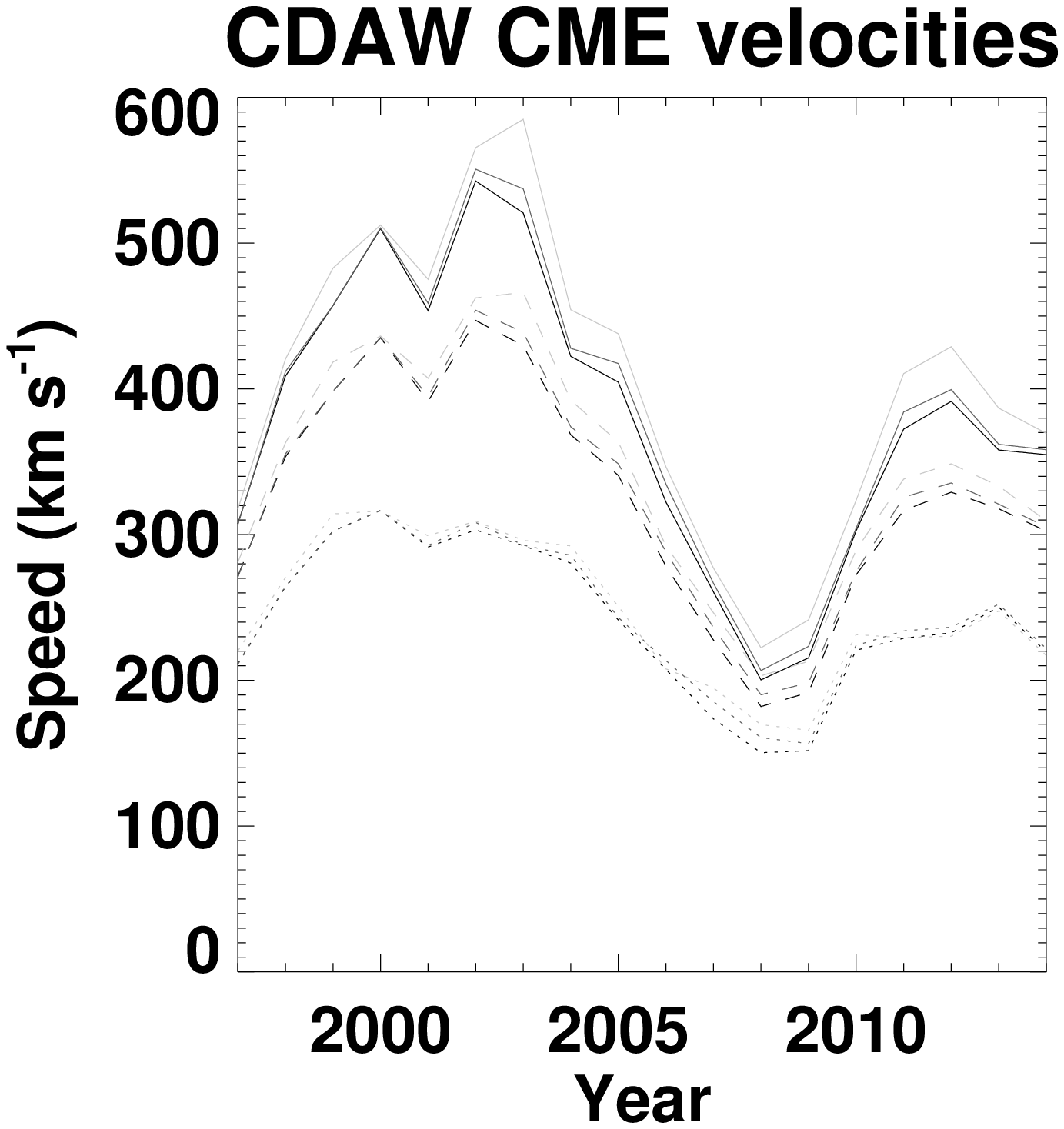}}
\resizebox{0.45\hsize}{!}{\includegraphics*{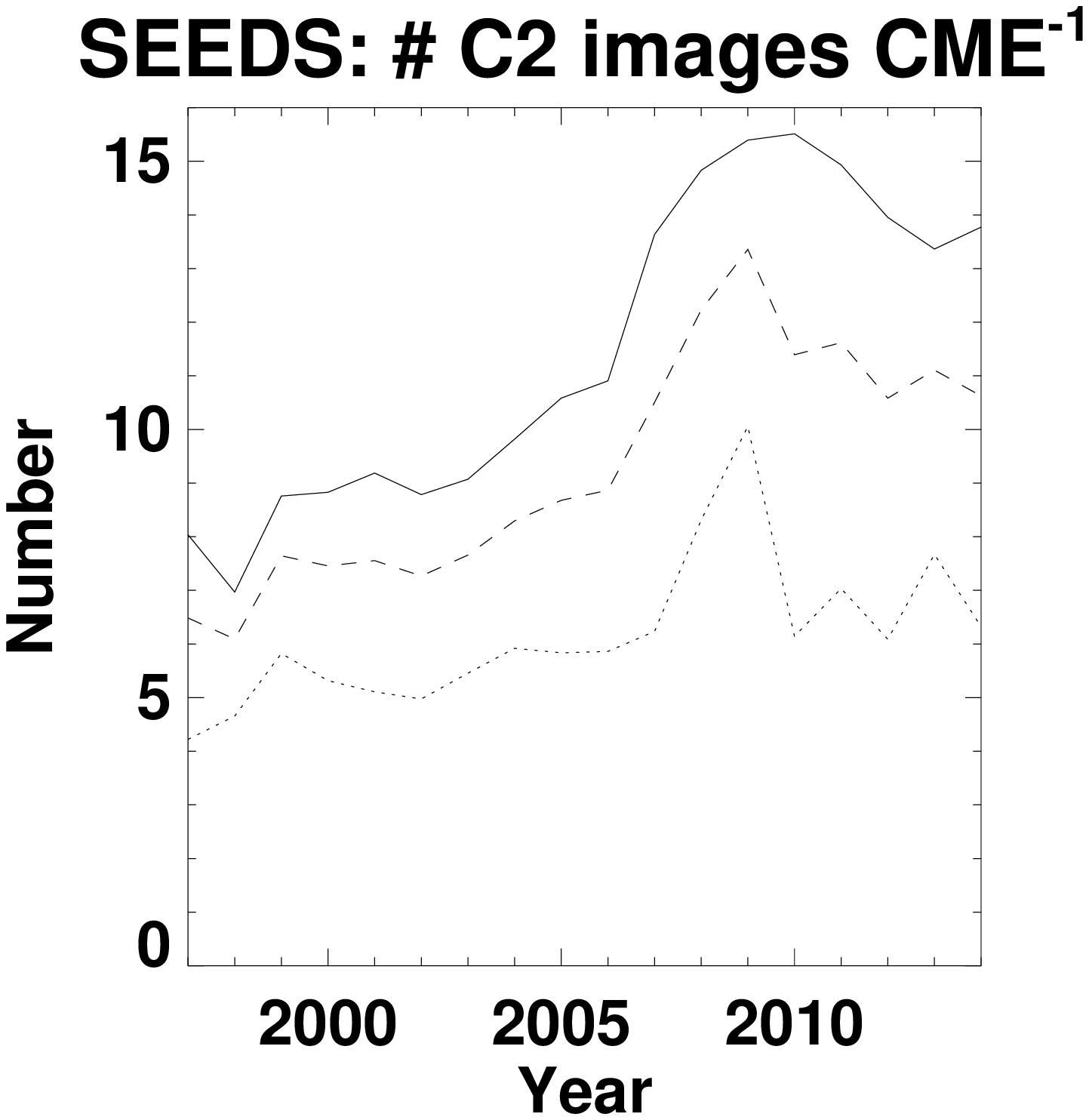}}
\end{center}
\caption{The annual mean (solid lines), median (dashed lines) and mode (dotted lines) of the log-normal velocity distributions in Figures~\ref{fig:velhists_seeds}-\ref{fig:velhists_cdawnop} are plotted against time for SEEDS (top left), CACTUS (top right) and CDAW (bottom left) detections. Here cases with angular width $< 30^{\circ}$ have been excluded from all databases. For CDAW data, the different shades of the curves represent calculations including ``poor'' and ``very poor'' detections (black), excluding ``very poor'' detections (dark grey) and excluding both ``poor'' and ``very poor'' detections (light grey). The bottom right plot shows the annual mean (solid lines), median (dashed lines) and mode (dotted lines) of the log-normal distributions of the annual SEEDS statistics for the number of LASCO C2 images per detected CME as functions of time.}
\label{fig:fitparams}
\end{figure}

It is apparent from Figures~\ref{fig:2dhists} and \ref{fig:cd2dhists} that the maximum speed of detected CMEs did not significantly increase as a result of the cadence change, nor is there an obvious increase in the median velocity. We can address the question of the median and mean velocities more qualitatively using Figure~\ref{fig:fitparams}, where the profiles of the modes, medians and means of the log-normal best fits to the velocity distributions are plotted. These best-fit log-normal functions are derived by optimizing the free parameters of Equation~(\ref{eq:lognormal}). Narrow CMEs (angular width $< 30^{\circ}$) are excluded from these calculations. The resulting time-profiles for the modes, medians and means of the log-normal best-fits shown in Figure~\ref{fig:fitparams} are strikingly similar for the CDAW and CACTus data, but somewhat different for the SEEDS data. The resemblance of the CDAW and CACTus plots to each other is likely due to the fact that the CDAW and CACTus databases are both based on C2 and C3 data, and the removal of the narrow CME detections has eliminated most of the effects of subjective CME identification in the CDAW database. The CDAW plot shows three sets of parameter profiles for CME detections with angular width $>30^{\circ}$: all such detections including ``poor'' and ``very poor'' detections (black lines), with ``poor'' detections excluded (dark grey lines), and with ``very poor'' detections excluded (light grey lines). The exclusion of ``poor'' and ``very poor'' detections does not produce a dramatic effect on the log-normal velocity distributions, but it increases the mean velocity by up to 10\% during the late maximum and decline of cycle 23 and cycle 24. This change is much smaller than the shift from faster average velocities of about 550~km~s$^{-1}$ at the peak of cycle 23 to a value of about 400~km~s$^{-1}$ typical for cycle 24 so far. A similar difference between cycle 23 and 24 peaks is seen in the CACTus data supporting the notion that a real physical change in velocity occurred between the two cycles. However, it appears that the enhanced cycle 23 peak only occurred in the two databases that extended to the C3 field of view: CACTUS and CDAW. The SEEDS mean and median velocities have only modest peaks in 2005, as shown in Figure~\ref{fig:fitparams}. Otherwise the SEEDS parameters maintain steady values throughout the active years of cycles 23 and 24. In particular, there is no obvious change in the velocity distributions between late cycle 23 (2004, 2005) and early cycle 24 (2011-2014).

The mode, median and mean values associated with the log-normal fits to the SEEDS statistics for C2 images CME$^{-1}$ (Figure~\ref{fig:nimhists}, left) are shown in the bottom right plot of Figure~\ref{fig:fitparams}. The mean and median C2 images CME$^{-1}$ show a clear stepwise change coinciding with the C2 image cadence change in 2010. The SEEDS velocity parameters do not show such a stepwise change. The CDAW and CACTus velocity parameters also have similar values before (2004, 2005) and after (2011-2014) the cycle 23 minimum. The agreement of all three databases on this point suggests that the LASCO image cadence change did not significantly impact the velocity distributions of the CMEs represented in Figure~\ref{fig:fitparams}, those with angular width $>30^{\circ}$.

\subsection{Comparison of CME rates, and the effects of excluding narrow and poorly detected CMEs}

\begin{figure}[h]
\begin{center}
\resizebox{0.9\hsize}{!}{\includegraphics*{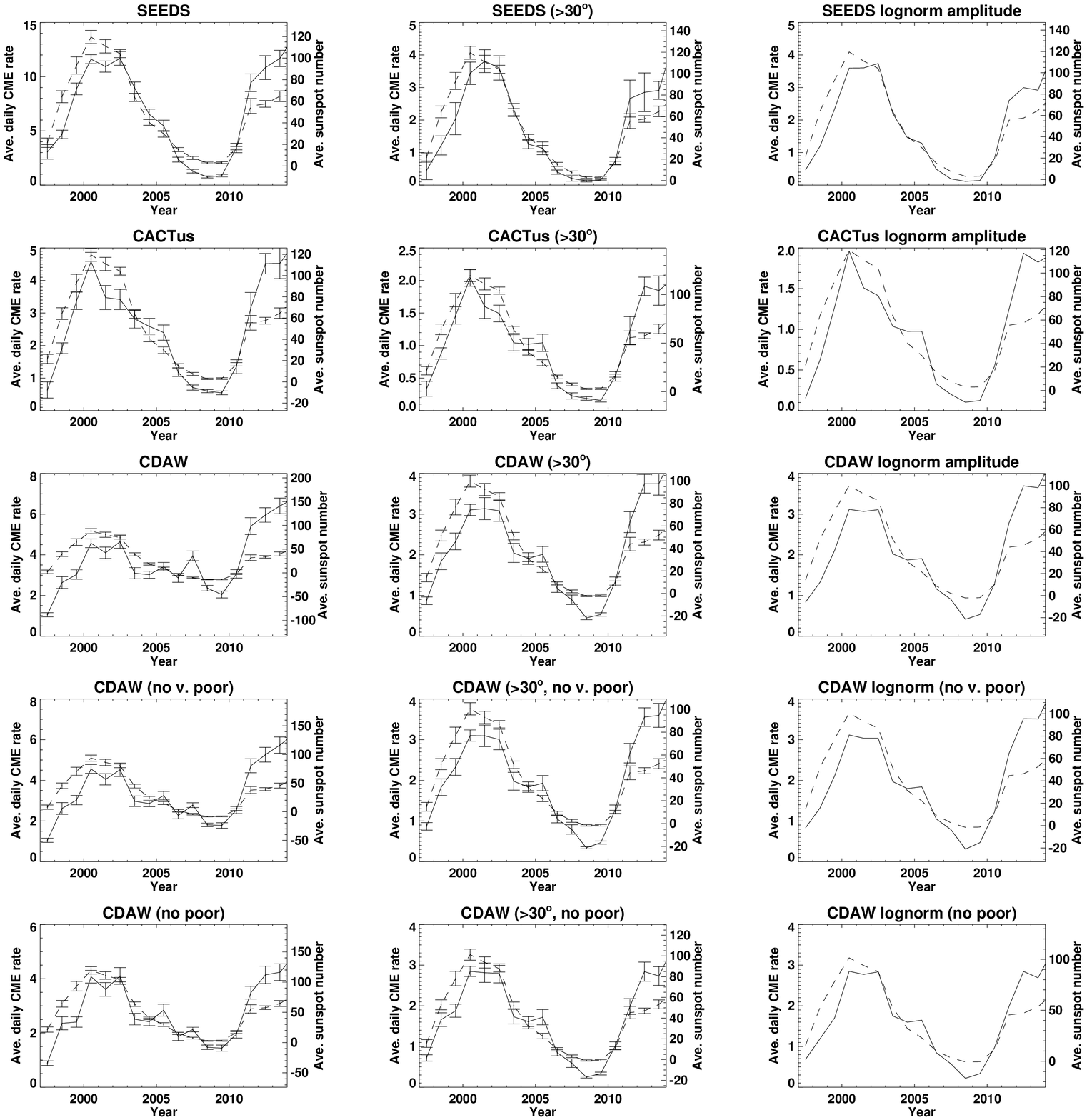}}
\end{center}
\caption{Average daily CME rates from SEEDS (top row), CACTus (second row) and CDAW (3rd-5th rows) with all cases included (3rd row), very poor cases excluded (4th row) and poor and very poor cases excluded (5th row). In the first column, CMEs of all angular widths are included, whereas in the middle column only CMEs with angular width $>30^{\circ}$ are included. The error bars indicate the standard deviations of the annual means. The right column shows the amplitude of Equation~(\ref{eq:lognormal}) fitted to the annual distribution of detections with angular width $>30^{\circ}$. The average monthly sunspot number is overplotted in dashed curves for comparison.}
\label{fig:cmessn}
\end{figure}

\begin{figure}[h]
\begin{center}
\resizebox{0.9\hsize}{!}{\includegraphics*{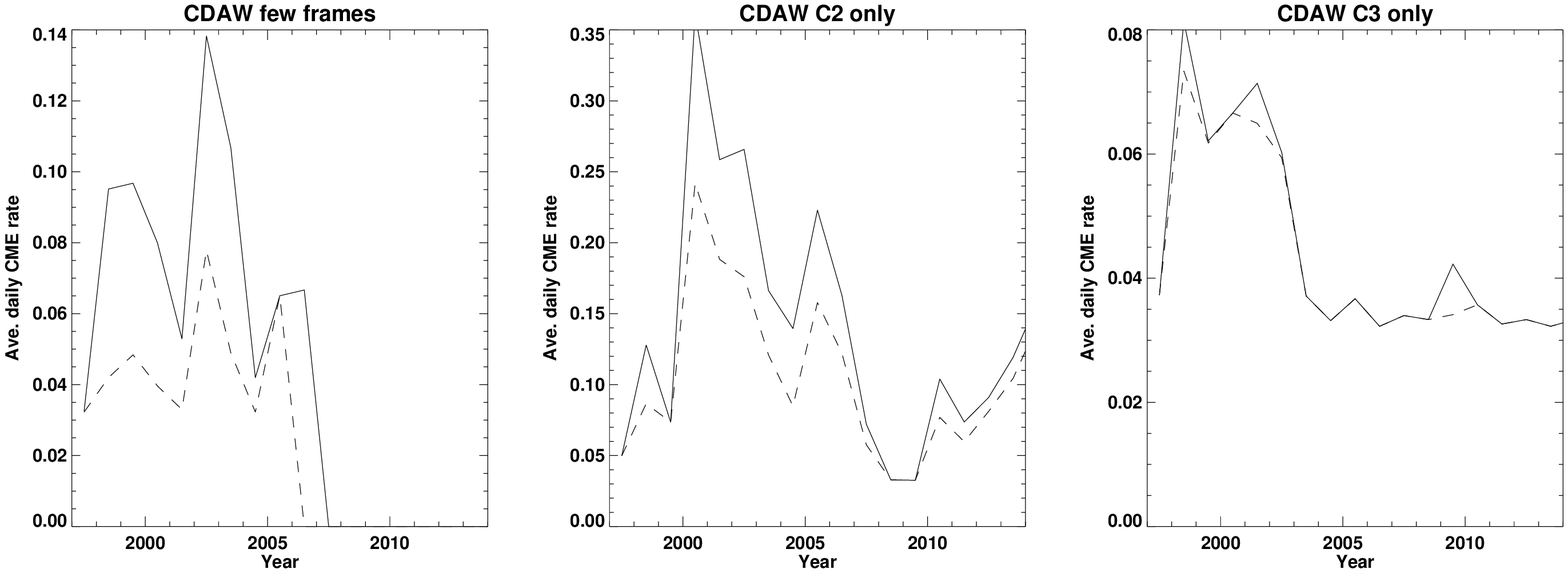}}
\end{center}
\caption{Average daily CME rates from CDAW where the CME appeared in only 2 or 3 LASCO images (left), only in C2 images (middle), and only in C3 images (right). Poor and very poor events are not included. Solid lines show the rates for all CMEs and dashed lines rates for CMEs with angular width $>30^{\circ}$.}
\label{fig:ocff}
\end{figure}

Figure~\ref{fig:cmessn} shows the annual CME rates for the SEEDS, CACTus and CDAW databases, including all detections (left plots) and for detections with angular width $>30^{\circ}$ (middle plots). Also shown are the amplitudes of the log-normal function fits for detections with angular width $>30^{\circ}$ (right plots). The agreement between the middle and right plots indicates that the log-normal functions represent the observed velocity distributions satisfactorily. In each plot the annual average monthly sunspot number is over-plotted, scaled according to the linear regression coefficients, and standard deviations of annual means are indicated both for CME rates and the sunspot number. This allows us to track the deviation of the CME rates from the sunspot number. Whereas in the past the CME rate has been well correlated with the sunspot number \citep[e.g., during cycle 21][]{WebbHoward94}, it is clear from Figure~\ref{fig:cmessn} that these LASCO-based CME rates deviate from the sunspot numbers in a statistically significant manner. In all cases the CME rate is systematically higher per sunspot number during cycle 24 than during cycle 23. On the other hand, the CME rates from the different databases differ significantly from each other in general. The CME detection rates evidently depend on differences in how the databases are constructed, including whether manual or automated detection methods were used, and whether the detections were based on C2 and/or C3 images.

As we saw in Figure~\ref{fig:cd2dhists}, Figure~\ref{fig:cmessn} also shows that the manually-derived CDAW statistics are affected by the inclusion of narrow CMEs in 2005 and some observer-dependent artifacts, giving the cycle 23 minimum disproportionately numerous detections and making the cycle 24 maximum peak much higher than the cycle 23 maximum. Again we see here, comparing the CDAW plots to the CACTus plots, that these artifacts are largely removed by excluding the CMEs with angular width $<30^{\circ}$.

Also shown in Figure~\ref{fig:cmessn} are CDAW CME rates excluding ``poor'' and ``very poor'' detections. The effects of these exclusions are more modest, but they do dampen the cycle 24 maximum peak so that it is approximately the same height as the cycle 23 maximum. The cycle 23 and 24 maximum peaks are also approximately the same height in the CACTus plots, including or excluding the CMEs with angular width $<30^{\circ}$.

In the plots of the SEEDS CME rates in Figure~\ref{fig:cmessn}, the cycle 23 and 24 maxima are approximately the same height when all detections are included, but the exclusion of narrow (angular width $<30^{\circ}$) CMEs shrinks both peaks by more than a half, and the cycle 24 peak becomes about 10-20\% smaller than the cycle 23 peak. Nevertheless there is a surplus of CMEs relative to the sunspot number after the cycle 23 sunspot maximum (2000-2002) compared to before. This surplus is reduced by the exclusion of narrow CMEs, those with with angular width $<30^{\circ}$. However, the surplus of CMEs per sunspot number in cycle 24 compared to cycle 23 remains large and significant for CMEs with for angular width $>30^{\circ}$. This significant enhancement of CME detections per sunspot number for cycle 24 relative to cycle 23 is therefore common to the SEEDS, CACTus and CDAW data.

The CACTus and CDAW curves in Figure~\ref{fig:cmessn} for angular width $>30^{\circ}$ also show significant increases relative to the sunspot number during the decline of cycle 23, in 2004 and 2005. Since this surplus appears in the CACTus and CDAW data, it seems to be independent of the observer biases in the CDAW data. However, this increase is absent from the SEEDS plot in Figure~\ref{fig:cmessn} for angular widths $>30^{\circ}$. According to \citet{Lamyetal14}, ARTEMIS recorded an enhanced CME rate during the rise of cycle 24 but no significant enhancement during the decline of cycle 23. This would agree with the SEEDS $>30^{\circ}$ plot in Figure~\ref{fig:cmessn}. Such agreement may be related to the exclusive use of C2 images by both SEEDS and ARTEMIS. That this late cycle 23 surplus occurs in the CACTus and CDAW data but not so prominently in the SEEDS (or ARTEMIS) data suggests that these extra CMEs may have required C3 images for their detection.

We now look at the effect of the LASCO image cadence change on the marginal CDAW detections, and the impact of these marginal detections on the overall statistics. Figure~\ref{fig:ocff} shows the average daily rate of detections by CDAW where the CME appeared in only 2 or 3 LASCO images, or only in C2 images, or only in C3 images. These rates are plotted separately for the cases with all angular widths and excluding those whose angular widths exceed $30^{\circ}$. Poor and very poor events are excluded. All of the rates plotted in Figure~\ref{fig:ocff} are low, and are much reduced in cycle 24 compared to cycle 23. The rate of CMEs detected in only 2 or 3 images was about one per 10 days (one per 20 days for angular widths $>30^{\circ}$) during cycle 23, and fell to zero in cycle 24. C2-only detections occurred at a rate of about one every 3-4 days during cycle 23 (one every 5-6 days for angular widths $>30^{\circ}$) and fell to one per 10 days in cycle 24. C3-only detections occurred every 15 days or so on average during cycle 23, almost all with angular width $>30^{\circ}$, and fell to one every 30 days in cycle 24. The LASCO image cadence changes for cycle 24 therefore seem to have significantly affected these marginal CME detections. As we discussed in the context of Figure~\ref{fig:nimhists}, the main effect of LASCO image cadence change seems to have been to increase the number of images per CME detection and to decrease the number of marginal detections as shown in Figure~\ref{fig:ocff}. However, their impact on the total CME statistics in Figure~\ref{fig:cmessn} was evidently small.



\subsection{The cycle 23/24 CME rate increase, and the decrease in the solar and heliospheric magnetic field strength}

\begin{figure}[h]
\begin{center}
\resizebox{0.7\hsize}{!}{\includegraphics*{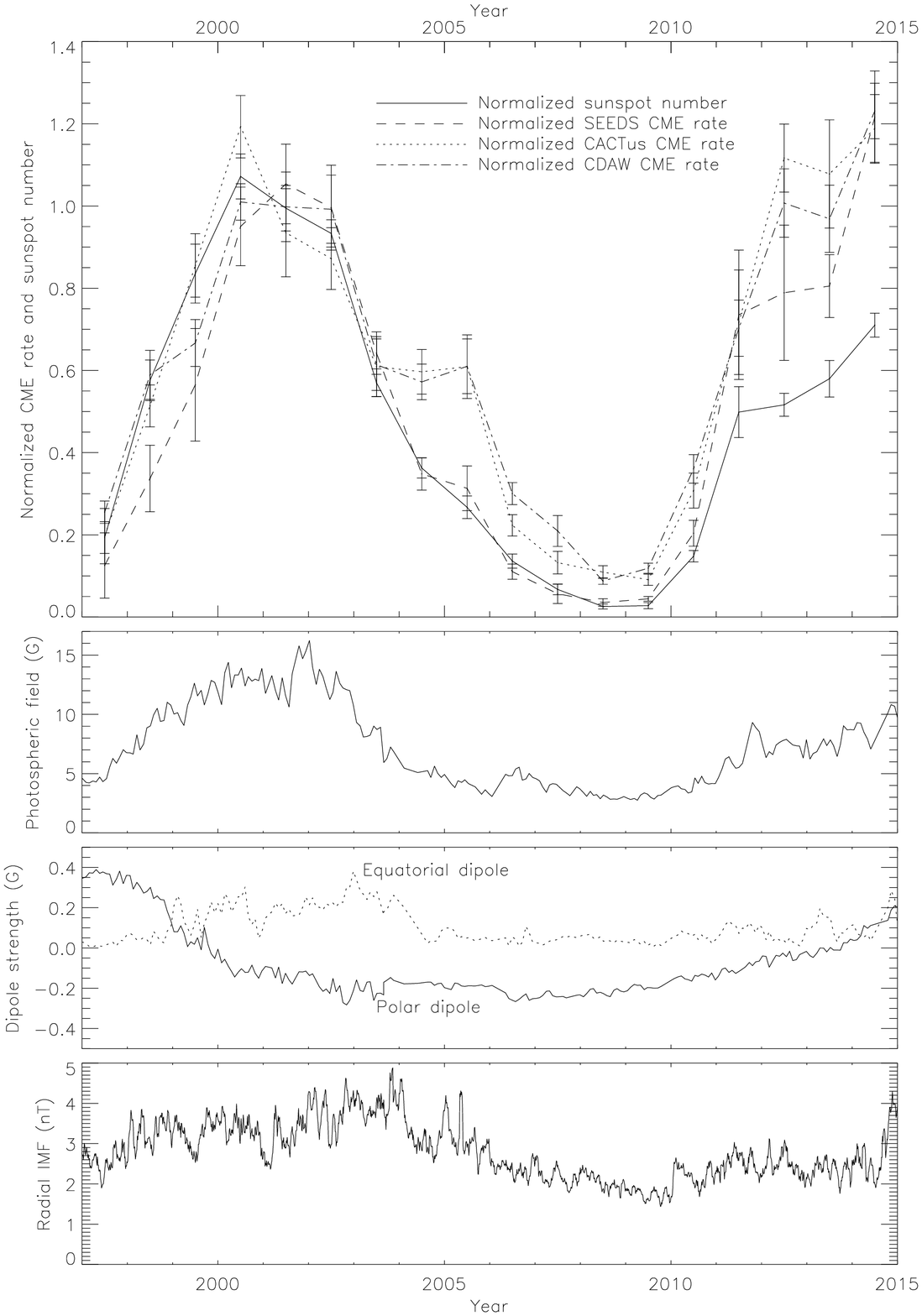}}
\end{center}
\caption{Normalized annual average daily CME rates (top panel) from SEEDS, CACTus  and CDAW including only detections with angular width $>30^{\circ}$, for the years 1997-2014. All poor and very poor cases have been excluded from the CDAW data. The normalized average monthly sunspot number is over-plotted in dashed curves for comparison. The CME rates have been normalized by their average values during the cycle 23 maximum years 2000-2002. Thus, the plot shows the divergence of their evolution since the cycle 23 maximum. The second panel shows the total photospheric magnetic flux from NSO KPVT and SOLIS/VSM synoptic magnetograms. The third panel shows the polar and equatorial dipole components of the same synoptic magneto grams. The bottom panel shows the OMNI 2 radial magnetic field component at 1~AU.}
\label{fig:cmemag}
\end{figure}

The top panel of Figure~\ref{fig:cmemag} shows a combined plot of the CME rates from SEEDS, CACTus and CDAW, for detections with angular width $>30^{\circ}$. Poor and very poor detections are excluded in the case of CDAW. The sunspot number is also over-plotted, and the CME rates and sunspot number are all normalized by their average values during the cycle 23 maximum years 2000-2002. Thus, the divergence of the CME rates from the sunspot number since the cycle 23 maximum is emphasized. In 2004, during the decline of cycle 23, the CACTus and CDAW CME rates diverged sharply from the sunspot number. Ever since 2004 these CME rates have remained elevated relative to the sunspot number by a statistically significant margin. Curiously, the SEEDS CME rate did not diverge from the sunspot number until the onset of cycle 23 in 2010-2011. The SEEDS CME rate caught up with the CACTus and CDAW rates in 2011, and has remained significantly elevated relative to the sunspot number throughout cycle 24 so far.

The second panel of Figure~\ref{fig:cmemag} plots the mean photospheric flux density from 1997 to 2015. The photospheric field is generally well correlated with the sunspot number, being dominated by active region flux during activity maxima. The photospheric flux has been around 40\% lower since the decline of cycle 23 compared to the ascent of cycle 23. The polar and equatorial dipole components of the photospheric field are plotted in the third panel of Figure~\ref{fig:cmemag}. The polar dipole represents the strength of the polar fields while the equatorial dipole follows the activity cycle. Both dipole components reflect the 40\% decrease in the photospheric field that occurred during the decline of cycle 23, around 2003-2004.

The fourth panel of Figure~\ref{fig:cmemag} plots the OMNI2 radial IMF component at 1~AU. The radial IMF has been widely reported to have decreased by about 30\% \citep{SmithBalogh08} between the cycle 22/23 and 23/24 minima, but the plot shows that this decrease actually occurred during the decline of cycle 23, early in 2004. During the cycle 23 declining phase, the photospheric polar fields ceased to strengthen around 2003, even though there was much magnetic activity between 2003 and the end of the cycle. \citet{Petrie12} argued that this was due to the the positive and negative active region flux latitude centroids statistically coinciding in each hemisphere, i.e., the Joy's law dipole tilt becoming insignificant, from around 2003 until the end of the cycle. This left the polar fields about 40\% weaker after their cycle 23 polarity reversal compared to before. This weakening of the photospheric field during the decline of cycle 23 produced a nearly stepwise 30\% decrease in the radial IMF in early 2004. On the rise of cycle 24 in 2010, the IMF only rose to cycle 23 declining-phase levels, about 30\% short of cycle 23 maximum-phase values. Recently the IMF has spiked as the polar fields have strengthened before the activity has decayed away.

It would be useful to relate the photospheric field to the IMF via a magnetic model for the coronal open flux. The obvious way to achieve this is the potential-field source-surface (PFSS) model \citep{Schattenetal69,AltschulerNewkirk69}. The solar open flux should match the radial component of the IMF integrated over the sphere at 1~AU. However, the PFSS model cannot be relied upon to model the open flux over full solar cycles with a fixed outer boundary (source surface) radius. Using the source surface radius as a free parameter, \citet{Leeetal09} showed that PFSS coronal hole maps and open fluxes better reproduced observed coronal hole distributions and IMF measurements with smaller source surface radii during the cycle 24 minimum than during the cycle 23 minimum, suggesting that the source-surface radius should be lowered when the photospheric field is weak. \citet{Ardenetal14} found that the source-surface radius ``breathes'', such that its height needs to be around 2.5 solar radii during solar minima and 15-30\% higher during solar maxima, according to comparisons between PFSS open fluxes and the measured IMF. This situation is further complicated by the general lack of quantitative agreement between photospheric magnetic field measurements from different observatories \citep{Rileyetal14}. We defer further investigation of the link between the photospheric and interplanetary fields to a future paper.

Comparing the panels of Figure~\ref{fig:cmemag}, the timing of the CACTus and CDAW CME rates' divergence from the sunspot number, in 2004, matches the decrease in the IMF. A causal physical link between the decreased IMF and an increased CME rate per sunspot number was suggested by \citet{Gopalswamyetal14}, who argued that a decreased heliospheric total (magnetic+plasma) pressure has allowed CMEs to expand more as they travel from the solar surface into the heliosphere, thereby enhancing the CME rate statistics.

The energy density of the radial magnetic field (that produces the transverse magnetic pressure stresses that would act upon an expanding CME) falls off much less rapidly than that of the transverse field \citep{Schattenetal69,Schatten71}. The energy density of the transverse field approximately balances the plasma pressure at about 0.6 solar radii above the photosphere. Thus the plasma extends the magnetic field outward near this point. In the case of the radial field, equality with the plasma energy density is only reached at the Alfv\'{e}n point near 25 solar radii \citep{Schatten71}. The magnetic field is therefore expected to constrain the plasma expansion out to to 25 solar radii, covering almost all of the fields of view of the LASCO C2 and C3 coronagraphs. Although \citet{Gopalswamyetal14} were comparing CME statistics from the rise phases of cycles 23 and 24, their argument links the CME rate to the solar and interplanetary magnetic field whose decrease in strength occurred during the decline of cycle 23, around 2003-2004. If the argument is correct then one would expect the increase in CME rate per sunspot number to begin around 2004, as the CACTus and CDAW statistics clearly do in the top panel of Figure~\ref{fig:cmemag}.

The question remains why the CACTus and CDAW CME rates diverged from the sunspot number in 2004, whereas the SEEDS rate waited until 2010-2011 before diverging. It is difficult to explain this in terms of a changing rate of CME expansion between the C2 and C3 fields of view. Such an explanation may again require improved coronal field modeling to estimate the magnetic (and total) pressure change between the C2 and C3 fields of view over time. It is worth noting, however, that the Nobeyama radio heliograph prominence eruption statistics also show a statistically significant in eruption rate per sunspot number from cycle 23 to cycle 24, and also a smaller but statistically significant increase per sunspot number around 2004 \citep{Petrie13}. Because eruptive prominences are almost always accompanied by CMEs \citep{Munroetal79}, prominence eruptions and CMEs can often be identified with each other \citep{Gopalswamyetal03b}, hence this increase in prominence eruptions around 2004 supports the notion that the CME rate also rose at these times.


\section{Conclusion}
\label{sect:conclusion}

The results are summarized as follows. In the three LASCO CME databases SEEDS, CACTus and CDAW, a statistically significant increase in the rate of CME detections with angular width $> 30^{\circ}$ per sunspot number was found for cycle 24 compared to cycle 23. In the two databases based on both LASCO/C2 and C3 images, CACTus and CDAW, the upward divergence of the CME rate relative to the sunspot number began in 2004 after the polar field reversal. At nearly the same time, the IMF decreased by $\approx 30\%$. These results are consistent with the the result of \citet{Gopalswamyetal14,Gopalswamyetal15} linking enhanced halo CME detections to increased CME expansion in a heliosphere of decreased total (magnetic+plasma) pressure. On the other hand, the SEEDS CME rate did not diverge from the sunspot number until the rise of cycle 24, in 2010-2011. It is possible that this can be explained by the restriction of the SEEDS detections to the C2 field of view. Such an explanation would need to rely on an improved model for the cycle dependence of the global coronal field strength at 2-3 solar radii than is currently available.

The LASCO C2 and C3 image cadence changes in 2010 may have some effect on the CME detection rates. However, this effect is likely to have been small in view of the increase in the number of images per detection, the decrease in low-quality detections and the lack of evidence of enhanced detections of very fast or faint CMEs only detectable in high-cadence sequences of images. Our restriction to angular widths $> 30^{\circ}$ focuses our study almost exclusively on CMEs that would be equally well detected using cadences of 10 or 20 minutes.
The increases in CME detection rate increases seem too large to be explained by marginal detection changes, and the explanation in terms of enhanced CME expansion in a coronal medium of reduced transverse magnetic pressure beginning in 2004 is consistent with the CACTus and CDAW data. The later increase of the SEEDS CME rate, at the beginning of cycle 24, remains to be explained.



\acknowledgements{}
The author thanks the referee for helpful comments. SOHO is a project of international cooperation between ESA and NASA. The CDAW CME catalog is generated and maintained at the CDAW Data Center by NASA and The Catholic University of America in cooperation with the Naval Research Laboratory.This paper uses data from the CACTus CME database, generated and maintained by the SIDC at the Royal Observatory of Belgium. The paper also uses CME statistics from the SEEDS project at George Mason University's Space Weather Laboratory, that has been supported by NASA's Living With a Star Program and NASA's Applied Information Systems Research Program. SOLIS data used here are produced cooperatively by NSF/NSO and NASA/LWS. NSO/Kitt Peak SPMG data used here were produced cooperatively by NSF/NOAO, NASA/GSFC, and NOAA/SEL. The OMNI data were obtained from the GSFC/SPDF OMNIWeb interface at http://omniweb.gsfc.nasa.gov

\end{document}